\documentclass[twocolumn,showpacs,preprintnumbers,amsmath,amssymb,superscriptaddress]{revtex4}
\usepackage{graphicx}
\begin{document}

\title{Photoemission study of poly(dA)-poly(dT) DNA : 
Experimental and theoretical approach to the electronic density of states}

\author{Hiroki Wadati}
\email{wadati@wyvern.phys.s.u-tokyo.ac.jp}
\affiliation{Department of Physics, University of Tokyo,
Bunkyo-ku, Tokyo 113-0033, Japan}

\author{Kozo Okazaki}
\affiliation{Department of Physics, University of Tokyo,
Bunkyo-ku, Tokyo 113-0033, Japan}

\author{Yasuhiro Niimi} 
\affiliation{Department of Physics, University of Tokyo,
Bunkyo-ku, Tokyo 113-0033, Japan}

\author{Atsushi Fujimori}
\affiliation{Department of Physics, University of Tokyo,
Bunkyo-ku, Tokyo 113-0033, Japan} 
\affiliation{Department of
Complexity Science and Engineering, University of Tokyo,
Bunkyo-ku, Tokyo 113-0033, Japan}

\author{Hitoshi Tabata}
\affiliation{The Institute of Scientific and Industrial
Research, Osaka University, 8-1 Mihogaoka, Ibaraki, Osaka, 567-0047,
Japan}

\author{Jared Pikus}
\affiliation{Department of Physics and Astronomy, Brigham
Young University, Provo, UT 84602-4658}

\author{James P. Lewis}
\affiliation{Department of Physics and Astronomy, Brigham
Young University, Provo, UT 84602-4658}

\date{\today}
\begin{abstract}
We present results of an ultraviolet photoemission spectroscopy study
of artificially synthesized poly(dA)-poly(dT) DNA molecules 
on $p$-type Si substrates. 
For comparison, we also present 
the electronic density of states (DOS) calculated using an \emph{ab
initio} tight-binding method based on density-functional theory (DFT). 
Good agreement was obtained between experiment and theory. 
The spectra of DNA networks on the Si substrate showed 
that the Fermi level of the substrate 
is located in the middle of the band gap of DNA. 
The spectra of thick ($\sim 70$ nm) DNA 
films showed a downward shift of 
$\sim 2$ eV compared to the network samples. 
\end{abstract}
\maketitle
There have been many proposals that DNA molecule is a promising material 
for the fabrication of molecular
electronic devices. The most striking application is to make excellent
one-dimensional wires. This proposal is especially
attractive, since advanced synthetic methods exist that produce, on-demand,
a wide variety of complex DNA sequences and structures. However, it
has been controversial whether DNA exhibits good conductivity or not. 
Also, it remains to be established whether the carriers are electrons or
holes. 
Eventually it is very
important for device applications to realize both $n$-type 
and $p$-type doping and desirable interfacial properties. 

The mechanisms 
of electron or hole transport in a DNA molecule have been intensively 
examined within the last few years. Pioneering experiments have produced
different interpretations and lively discussions 
\cite{Lewis976880,Fink99407,Porath00635,Cai00,Zhang02,Dandliker971466,
Beratan973,Berlin00443,
Voityuk00430,dePablo004992,Giese01318,Barnett01567,Yoo01198102,Hjort01228101,Yu016018}.
For example, Fink {\it et al.} \cite{Fink99407} observed a resistance on
the order of 1 M$\Omega $ for a 1-$\mu $m-long molecule (making
DNA a good conductor). 
In contrast, Porath {\it et al.} \cite{Porath00635}
observed a nonlinear $I-V$ curve with an insulating gap which makes DNA a
wide-gap semiconductor. 
Cai {\it et al.} \cite{Cai00} found that 
the resistance exponentially increases with the length 
of the DNA molecule in the atmospheric condition. 
Zhang {\it et al.} \cite{Zhang02} reported a high resistivity exceeding 
$10^6$ $\Omega$ cm. 
To address the above issues, photoemission spectroscopy is 
expected to provide unique and useful information. 
Also, the position of the Fermi level with respect to the 
band gap can be studied by photoemission spectroscopy. 
Recently, resonant photoemission spectroscopy  
near the Fermi level has been reported, indicating 
the localized unoccupied states of the bases \cite{Kato}. 
In the present work, we report on a photoemission study of 
artificially synthesized poly(dA)-poly(dT) in the film and network forms.

Poly(dA)-poly(dT) 
was synthesized by self-hybridization of poly(dA) of 50 mer (base
pairs) and poly(dT) of 50 mer 
(purchased from Amersharm Biosciences Co., Ltd).
The poly(dA)-poly(dT) samples were prepared as solutions
with a concentration of 1.25 mg/ml. In a previous study, we
found that the DNA 50 mers are assembled naturally and form a widely
spread network structure on SiO$_{2}$ surfaces with the concentration
of 1.25-0.25 mg/ml \cite{Tanaka01407}. It is known that isolated DNA molecules
yield deformed structures such as the relaxation of the pitch of the
double helix. In the case of the network, on the other hand, the original
molecule structure of the DNA molecules is maintained, that is,
the pitch of the helix remains about 3.4 nm, which corresponds to the pitch
of the ideal B-type form. 

To guarantee the structure of the DNA molecules with atomic level accuracy,
the molecules should be fixed on an atomically flat substrate. To
obtain the flat surface, we treated a Si(111) wafer with an H$_{2}$O$_{2}$-HCl
solution, an H$_{2}$O$_{2}$-NH$_{3}$ solution, an HF solution, and
a NH$_{4}$F solution sequentially as in the RCA method \cite{Yasuda}. 
Furthermore, we treated the H-terminated $p$-type 
Si(111) substrate with concentrated HNO$_{3}$ to obtain the SiO$_{2}$
surface. 

\begin{figure}
\begin{center}
\includegraphics[width=4cm]{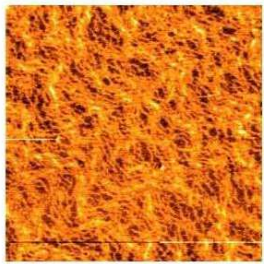}
\caption{Typical AFM image of the DNA network structure of 
poly(dA)-poly(dT). 
The scan area is 5 $\mu $m $\times $ 5 $\mu $m.}
\label{AFM}
\end{center}
\end{figure}

Both DNA network structures and DNA films (polycrystals 
covering the substrate) were then deposited
on the SiO$_{2}$ surface. To form the DNA network, the DNA solutions
in quantities of 20 $\mu $l were dropped on the SiO$_{2}$ surface and
blown off after 10 minute fixation. To form the DNA film, the DNA solutions
(also in quantities of 20 $\mu $l) were dropped on the SiO$_{2}$
surface and placed in vacuum for 3 or 4 hours until they dried completely.
The thickness of the film thus prepared was $\sim 70$ nm. 
The samples were characterized using an atomic force microscope (AFM)
in the tapping mode. A typical AFM image of the network structure
is shown in Fig.~\ref{AFM}. A DNA network structure was observed with
a height of 2.0-3.0 nm which suggests that the network was composed
of bundled DNA molecules. 

Photoemission measurements were performed using a VSW 
hemispherical analyzer and the He II resonance line 
($h\nu =40.8$ eV) of a VG helium discharge
lamp . 
The samples were introduced into the spectrometer through a fast-entry 
airlock. 
The energy resolution was set at about 100
meV. 
The Fermi level ($E_F$) position was determined using the spectra 
of gold which has electrical contact with the substrate. 
In order to remove absorbed water, we heated the samples to 350
K in the vacuum of the spectrometer, but no appreciable change was 
observed in the spectra. 
This suggests that there was 
negligible amount of absorbed water in our DNA samples. 
We also measured spectra of the Si substrate for background subtraction.

For comparison with the photoemission experimental results, the density of 
states (DOS) for 10 base pairs of poly(dA)-poly(dT) 
DNA sequences were calculated 
using first-principles methods. 
We obtained a series of 100 snapshots from
nanosecond classical molecular-dynamics simulations, separated by
0.5 ps (see Ref.~\cite{Lewis02} for details). A single-point calculation
of the DNA structure of each snapshot was performed using a local-orbital
method (called F{\scriptsize IREBALL}) based on density-functional
theory (DFT) and separable pseudopotentials \cite{Lewis01195103}.
For the purposes of these calculations, we choose a gradient corrected
exchange-correlation functional with Becke exchange and Lee-Yang-Parr
correlation, which is necessary for a hydrogen-bonded system such
as DNA. A double numerical basis set was used for H, C, N, O, and P.
Further details of these calculations are described in
Ref.~\cite{Lewis02} and Ref.~\cite{Lewis01195103}. 
The DOS for these 100 snapshots were averaged together to obtain the
average DOS; this averaging yields contributions
which include thermal and structural disorder to the electronic structure. 

\begin{figure}
\begin{center}
\includegraphics[width=6cm]{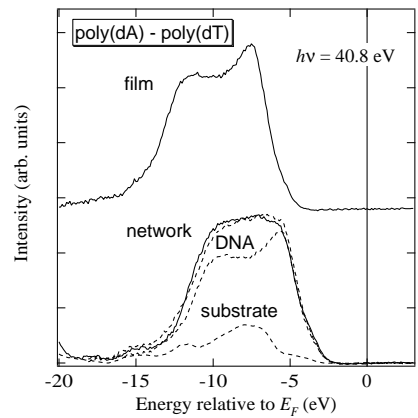}
\caption{Photoemission spectra of poly(dA)-poly(dT). 
Top : film sample. Bottom : network sample. 
The dashed curves are the film (shifted) and
 substrate spectra and their superposition. 
The solid curves are raw data.}
\label{UPS}
\end{center}
\end{figure}

Figure \ref{UPS} shows photoemission spectra of the poly(dA)-poly(dT) 
film and network. The spectra have been 
normalized to the area after the secondary
electron background subtraction using the procedure of 
Ref.~\cite{Li}. 
The spectrum shows a prominent emission band in the binding energy range 
$5-15$ eV for the film and $3-13$ eV for the network. 
The spectrum of the film consists of distinct 
two peaks with the lower binding energy peak being more pronounced. 
In the case of the network, the spectrum starts at about 2.5 eV below 
$E_F$ whereas in the film it starts at about 5 eV below $E_F$. 
One can see that the line shape of the spectrum of the network is different
from that of the film and is shifted toward a lower binding energy compared 
to that of the film. 

In order to see whether there was charging effect in the film sample, 
we took the spectra by changing the photon flux, and could not find
any energy shift. However, we cannot completely rule out 
the possibility of a charging effect, 
and more careful studies are necessary to establish 
the $E_F$ position in the film sample.
As for the line shape, We consider that 
the spectrum of the film represents the intrinsic
spectrum and that the spectrum of the network is a superposition 
of the intrinsic spectra of DNA and the spectrum of the substrate 
without the DNA deposition. 

\begin{figure}
\begin{center}
\includegraphics[width=6cm]{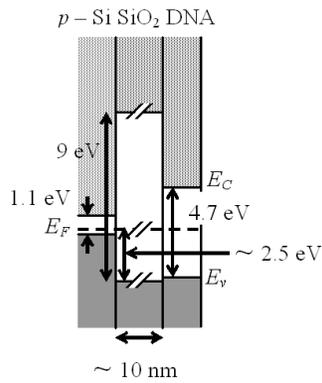}
\caption{Schematic band diagram of poly(dA)-poly(dT) 
DNA on the Si/SiO$_2$ substrate.}
\label{band}
\end{center}
\end{figure}

As shown in Fig.~\ref{UPS}, we could indeed reproduce 
the spectrum of 
the network from that of the film and that of the
substrate by shifting the film spectrum by $\sim 2$ eV and 
superimposing the substrate spectrum. 
In the visible ultraviolet absorption spectra, the absorption 
edge occurs at 4.7 eV for poly(dA)-poly(dT) \cite{Tabata}. 
Therefore, in the networks, where little charging effect is expected,  
the Fermi level is located in the middle of the band gap.
In fact, the spectrum of the Si/SiO$_2$ substrate is considered to 
be free from charging effect, in 
comparison with the previously reported spectra \cite{Spicer}. 
From these results, we obtained the schematic band diagram of 
poly(dA)-poly(dT) DNA on the Si/SiO$_2$ substrate as shown in 
Fig.~\ref{band}. (Possible band bending in the Si substrate 
is not taken into account.)

\begin{figure}
\includegraphics[width=6cm]{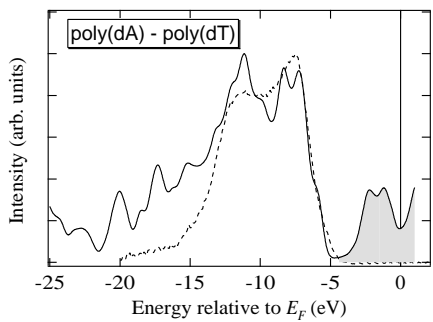}
\caption{Comparison of the theoretical DOS (solid curves)
 and the photoemission spectra (dashed curves) of poly(dA)-poly(dT). 
The hatched area corresponds to the unoccupied
 part of the calculated DOS.}
\label{calc} 
\end{figure}

Figure \ref{calc} shows comparison of the spectrum of the 
poly(dA)-poly(dT) film with the calculation. 
The calculated DOS has been broadened with a Gaussian of FWHM 
250 meV. 
The hatched area corresponds to the unoccupied part of the calculated 
DOS. 
Agreement between experiment and calculation is quite good. 
In comparison with the results of the calculation \cite{Lewis02}, 
the first peak ($\sim -7.5$ eV from $E_F$) is considered to 
be dominated by Adenine and Thymine and the second peak 
($\sim -12$ eV from $E_F$) by Ribose and Phosphate. 
The magnitude of the calculated band gap 
($\sim$ 2.5 eV for poly(dA)-poly(dT)) is much smaller than the 
experimental one of $\sim$ 4.7 eV,  
due to the well-known drawback of 
DFT to underestimate the band gap. 

Finally, we point out that the $E_F$ position of 
the Si substrate, which is located in the middle 
of the band gap of DNA, means that 
a bias of $\sim 2$ eV must be applied 
for the injection of carriers into the DNA networks 
on the $p$-type Si substrate. 
On the other hand, carriers created in the DNA 
molecules can be easily injected into the Si substrate. 
Since the spectra are shifted between the film 
and the network samples, it would be interesting 
to see how the $E_F$ position is shifted
as the thickness of the film increases.

In conclusion, we have measured the photoemission spectra of
poly(dA)-(dT) film and network. 
The spectrum of the film sample is considered to 
represent the intrinsic DOS of DNA. 
The position of the Fermi level in the network 
sample is located in the middle of the band gap. 
Overall, we obtain nice agreement between the experimental DOS 
and the DOS from DFT calculation except
for the magnitude of the band gap. 

We would like to thank Shin-ichi Tanaka for sample preparation and useful
discussions. 
We would also like to thank the following people for their enlightening
discussions regarding this ongoing project: M. Furukawa, 
M. Kawai, T. Kawai, O. Sankey, and H. Wang. 
Computer time from the Ira and Marylou Fulton Supercomputing Center 
(at BYU) was used in this work.

\bibliographystyle{prsty}
\bibliography{DNA1tex}

\end{document}